\documentclass[sigconf, nonacm, dvipsnames]{acmart} %
\AtBeginDocument{%
  \providecommand\BibTeX{{%
    \normalfont B\kern-0.5em{\scshape i\kern-0.25em b}\kern-0.8em\TeX}}}

\acmConference[Woodstock '18]{Woodstock '18: ACM Symposium on Neural
  Gaze Detection}{June 03--05, 2018}{Woodstock, NY}

\setcopyright{none} 
\usepackage{bbm}
\usepackage{tikz}
\usetikzlibrary{tikzmark}
\usetikzlibrary{calc}
\usepackage{bbm}
\usepackage{cleveref}
\usepackage[linesnumbered, vlined]{algorithm2e}
\usepackage{subcaption}
\usepackage[normalem]{ulem}
\usepackage{float}
\usepackage{pifont}

\newcommand{\cmark}{\ding{51}}%
\newcommand{\xmark}{\ding{55}}%

\newcommand*{\lmulti}{\{\mskip-5mu\{}
\newcommand*{\rmulti}{\}\mskip-5mu\}}

\newif\ifIEEE
\IEEEfalse
\newif\ifnotIEEE
\ifIEEE\notIEEEfalse\else\notIEEEtrue\fi

\newif\iflong
\longtrue

\newcommand{\panicc}{NeSt}
\newcommand{\mnf}{\ensuremath{\mathcal{N}}}

\newcommand\norm[1]{\lVert#1\rVert_1}
\newtheorem{claim}{Claim}
\DeclareMathOperator*{\argmax}{argmax}
\begin{document}

\title[\panicc{} Configuration Models]{Neighborhood Structure Configuration Models}%

\author{Felix I. Stamm}
\authornote{Both authors contributed equally to this research.}
\email{felix.stamm@cssh.rwth-aachen.de}
\affiliation{%
  \institution{RWTH Aachen University}
  \country{Germany}
}

\author{Michael Scholkemper}
\authornotemark[1]
\email{scholkemper@cs.rwth-aachen.de}
\affiliation{%
  \institution{RWTH Aachen University}
  \country{Germany}
}

\author{Markus Strohmaier}
\email{markus.strohmaier@uni-mannheim.de}
\affiliation{%
  \institution{University of Mannheim}
  \country{Germany}
}

\author{Michael T. Schaub}
\email{schaub@cs.rwth-aachen.de}
\affiliation{%
  \institution{RWTH Aachen University}
  \country{Germany}
  }

\renewcommand{\shortauthors}{F. Stamm and M. Scholkemper, et al.}

\begin{abstract}
We develop a new method to efficiently sample synthetic networks that preserve the $d$-hop neighborhood structure of a given network for any given $d$.
The proposed algorithm trades off the diversity in network samples against the depth of the neighborhood structure that is preserved.
Our key innovation is to employ a colored Configuration Model with colors derived from iterations of the so-called Color Refinement algorithm.
We prove that with increasing iterations the preserved structural information increases: the generated synthetic networks and the original network become more and more similar, and are eventually indistinguishable in terms of centrality measures such as PageRank, HITS, Katz centrality and eigenvector centrality.
Our work enables to efficiently generate samples with a precisely controlled similarity to the original network, especially for large networks.

\end{abstract}

\begin{teaserfigure}
  \includegraphics[width=\textwidth]{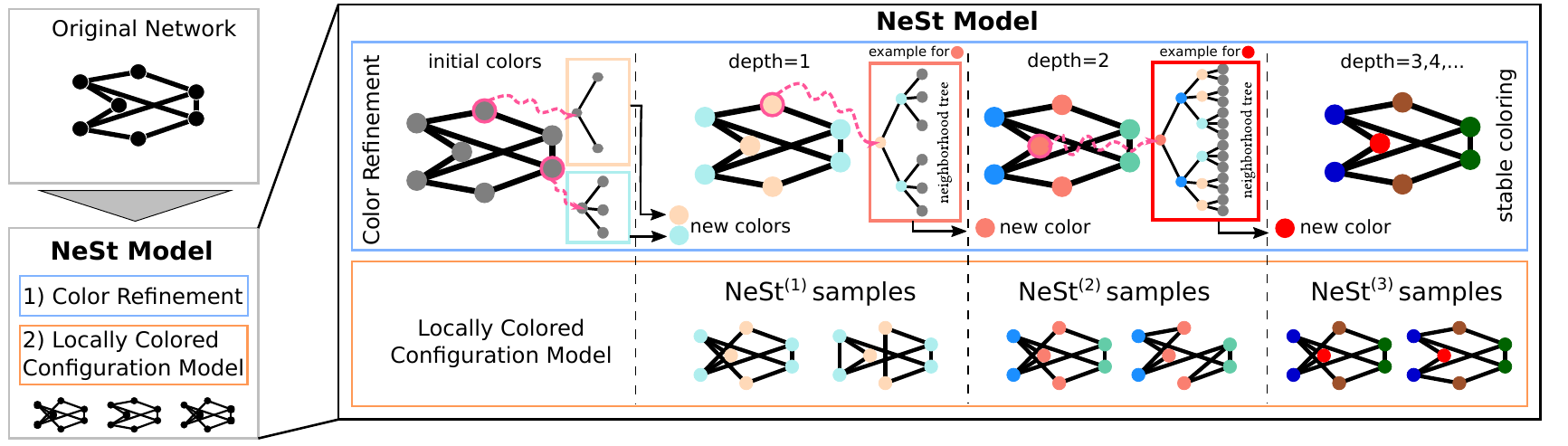}
\caption{
\protect Network sampling with the \panicc{} model.
Starting from an original network we use the Color Refinement (CR) algorithm to extract node colors for different depths $d$ (blue box).
At each depth $d$, nodes with an isomorphic neighborhood tree of depth $d$ are assigned the same color.
We then sample graphs via the locally Colored Configuration Model such that the CR colors of the original network are preserved. 
This ultimately preserves neighborhood trees as well.

}
  \label{fig:teaser}
\end{teaserfigure}

\maketitle

\section{Introduction}
Sampling networks with predefined properties is a fundamental problem in network science.
For instance, in order to assess whether a specific network statistic is significant, we need to compare its value against a baseline value, i.e., a reference statistic computed from a ``typical'' network drawn from some null model.
While a plethora of models exist, many models suffer from on of the following deficiencies:
On the one hand there exist null models that are efficient to sample from, but merely preserve local information about nodes such as their degrees or simple global network statistics such as the overall link density (examples include Erdos-Renyi random graph models, or configuration models).
Yet, in many applications preserving the meso-scale neighborhood structure of nodes is necessary for null models to resemble a given network more closely. 
On the other hand, null models that can preserve certain mesoscale structures and motifs, such as exponential random graph model, are notoriously difficult to fit to data and difficulty to sample, especially for larger networks.

In this paper, we address this research gap by introducing the \panicc{} models that both preserve mesoscale network structure and can be efficiently sampled.
More specifically, our models preserves the neighborhood trees around each node up to a specific depth $d$, and even for large networks are easy to fit, can be efficiently sampled, and well-approximates a given network on a range of centrality measures.
We achieve this by combining the so-called Color Refinement or Weisfeiler-Lehman algorithm~\cite{WL} (an approximate graph isomorphism test) with a locally Colored Configuration Model~\cite{CC}.
With the depth parameter $d$, we can tune how deep the multi-hop neighborhood structure of each node in the original network is preserved in the null model (See Fig. \ref{fig:teaser}  for an illustration).
Ultimately, certain spectral properties of the original network are \emph{exactly} preserved in samples from our network model.
We demonstrate the utility of \panicc{} by generating null networks which better and better preserve `spectral' centrality measures such as PageRank, Katz centrality, HITS, and eigenvector centrality with increasing depth.

\subsubsection*{Contributions}
We introduce a new class of network null models, the \panicc{}$_G^{d}(c^{(0)})$ model, whose samples mimic the original network $G$ in its neighborhood tree structure up to depth $d$ with starting colors $c^{(0)}$.
We further present an algorithm that allows efficient Markov Chain Monte Carlo sampling from the introduced model.
We prove that \panicc{} samples exactly preserve popular centrality measures of $G$ for an appropriate choice of $c^{(0)}$ and a large enough value of $d$.
For lower values of $d$, we show that early convergence of the PageRank power iteration in the original network carries over to corresponding \panicc{} samples.
Concluding, we illustrate empirically that similar low $d$ convergence observations can be made across a range of centralities and real-world networks.

\subsubsection*{Network null models}

There exists a broad range of network (null) models, including mechanistic models of network formation, models for growing networks, and many more. 
In this work we are concerned with \emph{null models} for static, fixed networks described by (potentially) directed graphs. 
The purpose of these network models is typically to preserve relevant properties (observables) of empirically measured real-world data, while otherwise maximizing randomness~\cite{cimini2019statistical, coolen2017generating}.
Depending on the chosen model, the desired network properties may either be preserved only in expectation, i.e., on average across the whole ensemble of generated networks or exactly in each sample.
We provide a non-exhaustive list  of some of the most commonly used null models in the following.

\noindent\textbf{Erd\H{o}s-R\'enyi type networks}
In Erd\H{o}s-R\'enyi (ER) random graphs, edges exist with equal probability, which preserves network density on expectation. Inhomogenous random graphs~\cite{soderberg2002general} (IRG) generalize ER-random graphs by allowing varying edge probabilities.
This enables the popular stochastic block model~\cite{SBM} (edge probability depends on group membership) and the Chung-Lu model~\cite{chung2002average} (degree distribution is preserved on expectation).

\noindent\textbf{Configuration models}
In the configuration model~\cite{newman2001random, fosdick2018configuring} (CM) the degree of each node is fixed in each sample.
In the globally colored configuration model~\cite{newman2003mixing} we associate a color to each node and fix the total number of edges between colors.
Alternatively, in locally-colored configuration models~\cite{CC} we fix the colors of all nodes' neighbors.

\noindent\textbf{Exponential Random Graph} models (ERGMs)~\cite{lusher2013exponential} are popular in the social sciences.
They specify structures to be preserved in expectation by including them in an ``energy term'' of a Gibbs distribution from which networks are sampled.
While the ERGM formulation is very flexible, fitting the parameters of the model is in general a difficult task~\cite{chatterjee2013estimating}, and sampling from these network models is often problematic~\cite{snijders2002markov}.

\noindent\textbf{Machine learning based network generators}~\cite{guo2020systematic}
 are gaining popularity in scenarios where multiple samples from one family of networks (e.g. enzymes) are available.
Deep learning methods like Variational Auto Encoders or Generative Adversarial Networks can then be employed to learn the distribution of this family of networks.
While these methods can learn arbitrary distributions, they are not easily employed when learning from large graphs.

\subsubsection*{Outline}
We start by introducing some prerequisites related to centrality measures and the color refinement algorithm.
Subsequently, we introduce the \panicc{} model, discuss how we can sample from this model and investigate its mathematical properties.
Finally, we show empirically how certain network centrality measures converge to those of the original network, even before the exact neighborhood tree structure of the original network is preserved (i.e., the final colors in the CR algorithm are reached).
We conclude the paper by highlighting limitations and potential further impact of our work.

\section{Preliminaries and Notation}
\label{sec:preliminaries}
\begingroup
\setlength{\thickmuskip}{1mu}
\textbf{Graphs.} A \textit{graph} or \textit{network} $G = (V,E)$ consists of a set of nodes $V$ and edges $E \subseteq \{uv \mid u,v \in V\}$. 
We always assume $V = \{1, ..., n\}$, thus graphs are \emph{labeled} and have an \emph{adjacency matrix} $A \in \{0,1\}^{n \times n}$ with $A_{i,j} = 1$ if $ij \in E$ and $0$ otherwise. 
We use parenthesis to distinguish matrix powers ($A^k$) from matrices indexed by superscripts ($A^{(k)}$).
A graph $G$ is \textit{undirected} if $uv \in E \Leftrightarrow vu \in E$, otherwise $G$ is \textit{directed}.
For directed graphs the \textit{in/out-neighborhood} is $N_{\text{in}/\text{out}}(v) = \{x \mid xv/vx \in E\}$ while for undirected graphs the \textit{neighborhood} $N(v) = N_{\text{in}}(v) = N_{\text{out}}(v)$. 
The degree $\text{deg}$ and in/out-degree $\text{deg}_{\text{in}/\text{out}}$ is the cardinality of the respective neighborhood sets.
\endgroup

\begingroup
\setlength{\thickmuskip}{1mu}
\noindent\textbf{Colorings.} A \textit{graph coloring} is a function $c : V \rightarrow \{1, ..., k\}$ that assigns each node one out of $k \in \mathbb{N}$  colors. 
Each coloring induces a partition $\mathcal{C}$ of the nodes into equivalence \textit{classes} of equal color $\mathcal{C}_i = \{v \in V \mid c(v) = i\}$. 
Given colorings $c_1, c_2$, we say $c_1$ refines $c_2$, denoted by $c_1 \sqsubseteq c_2$, if $c_1(v) = c_1(u) \Rightarrow c_2(v) = c_2(u)$. Similarly, if $c_1 \sqsubseteq c_2$ and $c_2 \sqsubseteq c_1$, $c_1$ and $c_2$ are \textit{equivalent}.
\endgroup

\noindent\textbf{Centrality measures.} Centrality measures assign importance scores to nodes such that important nodes have high centrality values.
In this work, we mostly consider \textit{eigenvector-based centralities} $\Gamma_{\textbf{X}}$, which compute the centrality scores of the nodes as the dominant eigenvector $w$ of certain matrices $M_X$, i.e.
$$
w, \text{ where } M_X w = \lambda_\text{max} w \text{ and } \lambda_\text{max} = \argmax_{\lambda_i \in \operatorname{spec}(M_X)} |\lambda_i|
$$
This is ill-defined when there are multiple dominant eigenvectors. We use a definition that ensures a unique centrality and agrees with the above if the dominant eigenvector is unique:
$$
\Gamma_{\textbf{X}} = \lim_{m \rightarrow \infty} \frac{1}{m} \sum_{i = 0}^m (|\lambda_{\text{max}}|^{-1} M_X)^i \mathbbm{1}
$$
This ensures unique centralities even when the graph is \emph{non primitive}.
For \textit{Eigenvector Centrality} \cite{bonacich1972technique}, the relevant matrix is
\begin{equation}
    \tag{$\Gamma_{\textbf{EV}}$}
    M_{\textbf{EV}} = A^\top .
\end{equation}
 
Similarly, the well-studied \textit{PageRank}~\cite{page1999pagerank} measure corresponds to the dominant eigenvector of the following matrix:
\begin{equation}
\tag{$\Gamma_{\textbf{PR}}$}
M_{\textbf{PR}} = \alpha \bar{A}^\top D^{-1} + (1-\alpha)\frac{1}{n} \mathbbm{1}_n \mathbbm{1}_n^\top,
\end{equation}
where $\alpha \in [0,1]$ is a so-called damping parameter and $\bar A$ is the adjacency matrix of an augmented graph in which we connect zero out-degree nodes to all nodes in the graph.
Thus the diagonal matrix of out-degrees $D = \operatorname{diag}(\bar A\mathbbm{1})$ is invertible.

\textit{HITS}~\cite{HITS} assigns each node both a hub-score $h_v$ and an authority score $a_v$. 
These are the dominant eigenvectors of:
\begin{equation}
    \tag{$\Gamma_{\textbf{hub}}$}
    M_{\textbf{auth}} = A^\top A \qquad \qquad (\Gamma_{\textbf{auth}}), \qquad M_{\textbf{hub}} = A A^\top
\end{equation}

Lastly, \textit{Katz centrality}\cite{katz1953new} is defined as:
\begin{equation}
    \tag{$\Gamma_{\textbf{Katz}}$}
    \Gamma_{\textbf{Katz}} = \sum_{k=0}^\infty \sum_{j = 1}^n a^k (A^k)_{j, \_} = \sum_{k=0}^\infty a^k (A^k)^\top \mathbbm{1}
\end{equation}
with $\frac{1}{a}>\max_{\lambda_i \in \operatorname{spec}(A)} |\lambda_i| $ being a parameter.%

\subsection{Color refinement}
\label{sec:wl_pr}

The color refinement algorithm (CR), also known as Weisfeiler Lehman algorithm and originally proposed in \cite{WL}, is a simple and efficient algorithm that is frequently used in the context of the graph isomorphism problem.
CR iteratively assigns colors to the nodes of the graph. Starting from an initial coloring $c^{(0)}$ the colors are updated by distinguishing nodes that have a different number of colored neighbors. CR stops once the color partition $\mathcal{C}$ no longer changes.
The initial coloring is typically chosen as constant over all nodes, but can also incorporate prior knowledge about the nodes provided.

The exact CR procedure follows the iteration:
\begin{equation}
c^{(d+1)}(v) = \operatorname{hash}\left(c^{(d)}(v),\;\lmulti c^{(d)}(x) \mid x \in N(v) \rmulti\right)
\label{eq:color_refinement}
\end{equation}
where the doubled brackets indicate a multi-set (a set in which an element can appear multiple times) and hash denotes some injective function mapping the pair onto a color space. This injectivity of the hash function implies that distinct inputs are assigned distinct colors.
Since the injective hash function takes the previous color as the first argument, the colorings are iteratively refined, i.e. $c^{(d+1)} \sqsubseteq c^{(d)}$. 
As there can be at most $n$ strict refinements $c^{(d+1)} \sqsubset c^{(d)}$, eventually the algorithm converges to a stable partition $c^{(d^*)} \equiv c^{(d^*+1)}$.

Once a stable partition $c^{(\infty)}$ is reached, the partition will not change.
At this point, the nodes' colors induce an \textit{equitable partition}, i.e., all nodes within one class have the same number of connections to another class. 
In fact, the CR algorithm converges to the \textit{coarsest} equitable partition of the graph \cite{paige1987three}.
As an example consider the graph in Figure~1. 
The partition at depth 3 is stable.
There all nodes of a specific color have the same number of colored neighbors as any other node of that color.
In contrast, the partition at depth 2 is not stable.
There the central red node has two blue neighbors while the top and bottom red nodes have one blue and one teal neighbor.

Though typically used for undirected graphs, the CR algorithm can be extended to directed graphs by replacing the neighborhood $N(v)$ in Eq.~\ref{eq:color_refinement} with either the in- or the out-neighborhood.
We refer to the resulting colorings as $c_\text{in}^{(d)}$ or $c_\text{out}^{(d)}$ respectively.
We may further distinguish nodes by both their in- and out-neighborhood:
\begin{equation}
\label{eq:color_refinement_both}
\begin{aligned}
c_{\text{both}}^{(d+1)}(v) = \operatorname{hash}\Big(c_{\text{both}}^{(d)}(v),\;& \lmulti c_{\text{both}}^{(d)}(x) \mid x \in N_{\text{in}}(v) \rmulti,\\
& \lmulti c_{\text{both}}^{(d)}(x) \mid x \in N_{\text{out}}(v) \rmulti \Big)
\end{aligned}
\end{equation}

Note that after $d$ iterations of the algorithm, the colors encode information about the $d$-hop (in-/out-)neighborhood of the nodes. To illustrate what we mean, once again consider \Cref{fig:teaser}, where we highlight a few \textit{neighborhood trees} with colored boxes. 
The neighborhood trees that correspond to the colors used at depth=1 are to the left of the depth=1 area. 
For instance, observe that the teal color corresponds to nodes that have three neighbors while the lightblue color corresponds to nodes that have two neighbors. 
Similar at depth two the lightred nodes correspond to nodes that have two neighbor both which have three neighbors. 
Overall, the color of a node directly encodes the structure of the neighborhood tree (up to a specific depth) in the sense that colors are the same if and only if their neighborhood trees are isomorphic \cite{barcelo2020logical}.

For directed graphs, the in- or out-neighborhood are isomorphic depending on the employed CR variant, e.g., for $c^{(\infty)}_{\text{both}}$ both in- and out-neighborhood are isomorphic, whereas for $c^{(\infty)}_{\text{in}}$ only in-neighborhood trees of the nodes are isomorphic.

The CR algorithm has a worst-case runtime of $\mathcal{O}((|E|+|V|)\cdot\log(|V|))$ \cite{grohe2014dimension}. 
However, we use a variant that has worst-case runtime $\mathcal{O}(d \cdot |V| \cdot \text{deg}_{\text{max}} \cdot \log(\text{deg}_{\text{max}}))$, which is preferable on most real-world graphs for which typically $d \ll |V|$ and we often only care about colorings corresponding to small $d$.

\section{The \panicc{} model}
\label{sec:model}

In this section we introduce the \textbf{Ne}ighborhood \textbf{St}ructure Configuration model, short the \panicc{} model. This model preserves neighborhood structure information of an original network up to a specified depth $d$ as encoded with the Color Refinement Algorithm. The locally Colored Configuration Model is then used to flexibly generate surrogate networks for a given network.
Importantly, due to its design, the \panicc{} model is computationally easy to fit and sample.

For a given labeled graph $G$ and initial node colors $c^{(0)}$, the set $\mnf_G^{d}(c^{(0)})$ contains all labeled graphs whose nodes have the same $d$-round CR colors as the original graph.
The \panicc{} model is the uniform probability distribution over $\mnf_G^{d}(c^{(0)})$ for $d \in \mathbb{N}^+$.
We think of initial colors $c^{(0)}$ and depth $d$ as the models' hyper parameters, while the remaining parameters are learned from the graph $G$.
Note that preserving the neighborhood tree structure at depth $d$ also preserves the neighborhood structure at depth $d' \leq d$, which implies that the sets of possible networks generated by the \panicc{} model are nested.

Before embarking on a more detailed technical discussion, we state here several noteworthy properties of the \panicc{} model:
\begin{enumerate}
    \item  $\mnf{}_G^{(1)}(\text{const})$ recovers the standard configuration model with degree sequence identical to $G$
    \item The graphs $G' \in \mnf{}_G^{(d)}(\text{const})$ and $G$ are identically colored during the first $d$ steps of the CR-algorithm
    \item The set $\mnf{}_G^{(d)}$ contains \emph{all} (simple) graphs that agree with $G$ on the first $d$ steps of the employed CR-algorithm
    \item Structure preserved in $\mnf{}_G^{(d)}$ is also preserved in $\mnf{}_G^{(d+1)}$
    \item $\mnf{}_G^{(d)}(c_0)\subseteq \mnf{}_G^{(d)}(const)$
\end{enumerate}
In the following we describe how we can efficiently sample from \panicc{}, and outline several variations to enrich the standard \panicc{} model and tailor it to specific scenarios.

\subsection{Sampling from the \panicc{} model}

In this section we outline how we can sample efficiently from the \panicc{} model to generate networks which preserve the neighborhood structure of the original network up to a specified depth $d$.

\begin{figure}[tb]
		\begin{subfigure}[b]{0.47\columnwidth}
			\includegraphics[width=\textwidth]{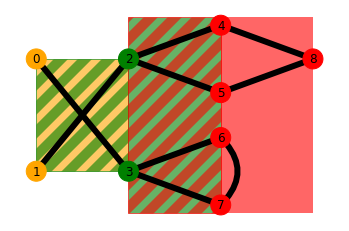}
			\caption{Original network}
		\end{subfigure}\hspace{2mm}
		\begin{subfigure}[b]{0.47\columnwidth}
			\includegraphics[width=\textwidth]{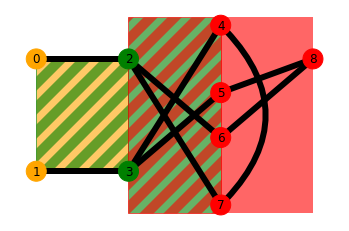}
			\caption{\panicc{} rewired network}
		\end{subfigure}\hspace{2mm}
		\caption{Illustration of the rewiring procedure employed when sampling from the \panicc{}$^{(2)}$ model. 
		First, the edges are partitioned into subgraphs according to the colors of depth $1$ of their endpoints. 
		The striped regions with two colors (yellow-green, green-red) correspond to bipartite networks connecting nodes of different colors while the unicolored regions (red) connect same color nodes.
		Second, we rewire each subgraph (colored regions) while respecting bipartivity as required.
		Overall this procedure preserves the CR-colors at depth $2$.}
	\label{fig::rewiring}
\end{figure}

To sample from the \panicc{} model we proceed as follows.
First, we partition the edges of the initial graph according to the colors of their endpoints into disjoint subgraphs $g_{C_i, C_j}$ with $V(g_{C_i, C_j}) = C_i \cup C_j$ and $E(g_{C_i, C_j}) = \{xy \mid x \in C_i, y\in C_j, xy \in E(G)\}$.
As an example, all edges connecting green and red nodes are put into $g_{\text{green}, \text{red}}$ while edges connecting red to red nodes are put into $g_{\text{red}, \text{red}}$ (see \cref{fig::rewiring} for such a partition).
In the case of directed networks, we distinguish $g_{\text{green}, \text{red}}$ and $g_{\text{red}, \text{green}}$ by the direction of the edge.

Second, after we have partitioned the edges into subgraphs, we can randomize each subgraph via edge swaps. In such an edge swap we choose two edges at random and swap their endpoints with one another.
A few points in this context are worth remarking upon.
For unicolored subgraphs all edge swaps that do not introduce multiedges are acceptable.
The subgraphs can thus be rewired as in the normal configuration model.
The subgraphs containing edges with endpoints of different color are bipartite.
To ensure a consistent neighborhood preserving sampling we can thus only allow edge swaps that preserve this bipartite structure.
These subgraphs are thus rewired as a bipartite configuration model.

For directed graphs a similar scheme can be used: The unicolored and two-color subgraphs are randomized according to the directed configuration model.
Unlike in the undirected case, the bipartivity of the two-color subgraphs is preserved automatically, since the edges are directed from $C_i$ to $C_j$, swapping endpoints always preserves the color at the end.
For unicolored directed configuration models an additional triangle swap is required for correct rewiring (see \cref{sec:appendix}).

The entire procedure is detailed in Algorithm~\ref{alg:sampling}. 
We note that all edges in distinct subgraphs are independent, and thus the sampling can run in parallel for different subgraphs.

As outlined, the sampling procedure can be decomposed into drawing samples from well established configuration models. We can thus draw upon the established research on sampling from these models via MCMC using random edge switches~\cite{artzy2005generating, erdos2019mixing}, which is the strategy we apply in this work,
However, other strategies are also possible~\cite{carstens2015proof}.
We assert the utility of the sampling procedure with the following theorem:

\begin{theorem}
\label{thm:sampling-correct}
The sampling procedures shown in algorithm~\ref{alg:sampling} and algorithm~\ref{alg:sampling2} sample uniformly at random from all labeled graphs in $\mnf^{d}_G(c^{(0)})$ for a given graph $G$, initial colors $c^{(0)}$ and depth $d$.
\end{theorem}

The proof can be found in the appendix (\cref{sec:appendix}). 
Note that the above result not only establishes the correctness of the algorithm but also that the samples drawn are \emph{maximally random} in the sense that we draw uniformly from the space of all graphs that respect the desired neighborhood constraints. This also means the \panicc{} model is the maximum entropy distribution over the set $\mnf_G^{d}(c^{(0)})$.

We briefly comment on the intuition that edge swaps of this fashion do not affect the colors obtained by CR.
To preserve the colors $c^{(d)}$ of a node from the original graph, it is sufficient to preserve the node color and the multiset of it's neighboring nodes' colors of the previous CR iteration ($c^{(d-1)}$).
By performing the outlined edge swaps, the degree of each node within the subgraphs stays the same. Thus for any color, the number of neighboring nodes with that color stays the same. Consequently, the multiset of neighboring nodes' colors stays the same.
Note that preserving the local configuration of neighborhood colors of each node is \emph{not} the same as preserving merely the total number of edges between color classes as done in other configuration models~\cite{newman2003mixing}.

\begin{algorithm}[t]
 \KwIn{Graph $G$, initial colors $c^{(0)}$, depth $d$}
 \KwOut{Sample Graph $G'$ distributed as \panicc{}$_G^d(c^{(0)})$}
 use CR to obtain depth $d-1$ colors $c^{(d-1)}$\\
 partition edges $ij$ of $G$ into subgraphs $g_{C^{(d-1)}_i, C^{(d-1)}_j}$\\
 let $\mathcal G$ be a list of all those subgraphs\\
 \For{subgraph $g \in \mathcal G$}{
    \For {$r \cdot |E(g)|$ steps}{
        \scalebox{0.89}[1.0]{choose two edges $u_1 v_1$ and $u_2 v_2$ unif. at random from $E(g)$}\\
        \uIf{$|\{u_1, v_1, u_2, v_2\}| = 4 $}{ 
            \uIf {$u_1 v_2 \notin E(g) \text{ and } u_2 v_1 \notin E(g)$}{
                remove $u_1 v_1, u_2 v_2$\\
                \tikzmarknode{anch}{a}dd $u_1 v_2, u_2 v_1$\\
            }
        }
        \uIf {\scalebox{0.92}[1.0]{$c^{(d-1)}(u_1) = c^{(d-1)}(v_1)$}}{%
 \begin{tikzpicture}[remember picture,overlay,scale=0.5,every node/.style={scale=0.5}]%
               \node (alpha) at ($(anch.west)+(7.2, 0.0)$)%
                 {%
                    \begin{tikzpicture}[scale=1.0, every node/.style={scale=1.0, inner sep=1pt, fill=black, circle, text=white, font=\huge}]
                \node[] (a) at (0,0) {$u_1$};
                \node[] (b) at (0,1.5) {$v_1$};
                \node[] (c) at (1.5,0) {$u_2$};
                \node[] (d) at (1.5,1.5) {$v_2$};
                \draw [->,line width=2pt] (a) to (b);
                \draw [->,line width=2pt] (c) to (d);
                    \end{tikzpicture}
                 };%
                \node (beta) at (alpha.east) [anchor=west,xshift=1cm]
                 {%
                    \begin{tikzpicture}[scale=1.0, every node/.style={scale=1.0, inner sep=1pt, fill=black, circle, text=white, font=\huge}]
                    \node[] (a) at (0,0) {$u_1$};
                    \node[] (b) at (0,1.5) {$v_1$};
                    \node[] (c) at (1.5,0) {$u_2$};
                    \node[] (d) at (1.5,1.5) {$v_2$};
                    \draw [->,line width=2pt] (a) to (d);
                    \draw [->,line width=2pt] (c) to (b);
                    \end{tikzpicture}
                 };%
               \draw [->,line width=1.0pt] (alpha)--(beta);%
\end{tikzpicture}%
            randomly choose three nodes $u_1, u_2, u_3$\\
            \uIf{$u_1, u_2, u_3$ constitute a directed triangle}{
                reverse direction of triangle $u_1, u_2, u_3$
                \begin{tikzpicture}[baseline=+0.2cm,scale=0.6,every node/.style={scale=0.6}]
               \node (alpha) at (0,0)
                 {
                    \begin{tikzpicture}[scale=0.6, every node/.style={scale=0.6, inner sep=1pt, fill=black, circle, text=white, font=\huge}]
                    \node[] (a) at (90:1) {$u_1$};
                    \node[] (b) at (210:1) {$u_2$};
                    \node[] (c) at (330:1) {$u_3$};
                    \draw [->,line width=2pt] (a) to (b);
                    \draw [->,line width=2pt] (b) to (c);
                    \draw [->,line width=2pt] (c) to (a); %
                    \end{tikzpicture}
                 };
                \node (beta) at (alpha.east) [anchor=west,xshift=1cm]
                 {
                    \begin{tikzpicture}[scale=0.6, every node/.style={scale=0.6, inner sep=1pt, fill=black, circle, text=white, font=\huge}]
                    \node[] (a) at (90:1) {$u_1$};
                    \node[] (b) at (210:1) {$u_2$};
                    \node[] (c) at (330:1) {$u_3$};
                    \draw [->,line width=2pt] (b) to (a);
                    \draw [->,line width=2pt] (c) to (b);
                    \draw [->,line width=2pt] (a) to (c); %
                    \end{tikzpicture}
                 };
               \draw [->,line width=1.0pt] (alpha)--(beta);
            \end{tikzpicture}
            }
        }
    }
 }
 return $G' = \bigcup_{g\in \mathcal G} g$%

 \caption{Sampling from \panicc{}$^{(d)}_G(c^{(0)})$ using edge switches.
 Firstly the CR colors are computed and edges are partitioned into smaller subgraphs according to the colors of their endpoints (lines 1-3).
 We then visit each subgraph in turn and perform a number of switching attempts proportional to the number of edges in the subgraph (lines 4-5).
 For undirected graphs simple edge switches (lines 6-10) are sufficient to achieve randomization. For the \emph{directed} case an additional directed triangle switch (lines 11-14) is required for uniform sampling (see the proof in \cref{sec:appendix}). This overall switching strategy (lines 6-14) is well established see \cite{artzy2005generating, erdos2019mixing}.
The runtime of the entire procedure is $\mathcal{O}(r \cdot |E(G)| \cdot \text{deg}_\text{max})$ - disregarding the computation time needed for the CR algorithm. The loop in line 4 is independent making the bulk of the computation easy to parallelize.
 }%
 \label{alg:sampling}%
\end{algorithm}

A different way to sample from the NeSt model is displayed in algorithm~\ref{alg:sampling2}. Here instead of rewiring each subgraph independent from another as in algorithm~\ref{alg:sampling} we are performing a randomisation one flip at a time. For each flip we choose a new graph $g$ in which we perform the flip. This makes it more similar to other MCMC algorithms such as those used when sampling from ERGM models.
The random choice of the subgraph $g$ in line 5 doesn't matter for correctness as long as all graphs have a non zero chance to be selected and that selection chance doesn't change throughout execution. But the choice could matter for the mixing time of the Markov chain i.e. the number of steps required to achieve approximately uniform sampling.

\begin{algorithm}[t]
 \KwIn{Graph $G$, initial colors $c^{(0)}$, depth $d$}
 \KwOut{Sample Graph $G'$ distributed as \panicc{}$_G^d(c^{(0)})$}
 use CR to obtain depth $d-1$ colors $c^{(d-1)}$\\
 partition edges $ij$ of $G$ into subgraphs $g_{C^{(d-1)}_i, C^{(d-1)}_j}$\\
 let $\mathcal G$ be a list of all those subgraphs\\
 \For{a number of steps}{
    randomly choose a subgraph $g$ from $\mathcal G$
    
   do lines 6-14 of Algorithm 1

 }
 Return $G' = \bigcup_{g\in \mathcal G} g$
 \caption{Alternative way to sample from \panicc{}$^{(d)}_G(c^{(0)})$. Instead of attempting a fixed number of switches in each subgraphs $g$ we allow for a variable number of switches by randomly choosing the subgraph to rewire (line 5). This makes sampling more similar to other MCMC sampling algorithms like those used for ERGMs and allows for a more fair comparison with those methods. We note that this algorithm is much more sequential than algorithm 1 as the loop in line 4 is no longer independent.
 }
 \label{alg:sampling2}
\end{algorithm}

\subsection{Variations of the \panicc{} model}
\label{sec::variants}

To emphasize the flexibility of our proposed scheme we discuss some variations of the \panicc{} model in this section.

\noindent\textbf{The initial node colors}
in the CR algorithm are a powerful way to incorporate additional constraints to the network ensemble. 
For concreteness, let us consider two simple examples.
\begin{itemize}
    \item Assume the original network is bipartite and we want to maintain this property. In this case, one can choose the initial colors to reflect the bipartition.
    \item Assume a network comprises several connected components we want to keep separated. In this case, we can choose the initial node colors to reflect the components.
\end{itemize}

A more elaborate use of the initial colors would be the following.
Assume we want to preserve the early steps of the PageRank power iteration.
One possible strategy to achieve this is to color our graph with the out-degree of the nodes and then perform the CR algorithm using the in-neighborhood.
In fact, this idea is formalized in \cref{lemma:intermediate_colors}.

\noindent\textbf{Incorporating externally defined node colors}
If external node colors are available, we can use them to initialize the CR algorithm.
This implies that the external colors are used throughout the CR process, i.e., the external-colors of nodes at any depth in the neighborhood are preserved.
This can be a strong restriction on the set of possible graphs in the \panicc{} models.
An alternative, less restrictive way to introduce external colors is to use them as function arguments in later iterations of the CR algorithm.
As an example consider injecting the external colors at iteration $d-1$ when sampling from \panicc{}$^{(d)}$.
In this case only direct node neighbors are additionally identified by their external color while two-hop neighbors are not identified by their external color.

\noindent\textbf{Samples in between depth $d$ and $d+1$}
For certain graphs, the cardinality of the set of possible surrogate samples can shrink drastically when moving from depth $d$ to depth $d+1$.
In those cases, it can be useful to preserve more neighborhood structure than depth $d$ but less than depth $d+1$.
To illustrate how this can be achieved, observe that the CR algorithm as described above, can alternatively be understood as a two step procedure: 
1) each node `pushes' its depth $d$ color to all its neighbors and itself; 
2) each node uses the so collected multi-set of colors to create the round $d+1$ color.
To retain the neighborhood structure in between $d$ and $d+1$, one can adjust the first pass in this alternative view of CR to only use all those nodes belonging to a selected (e.g., random) subset of depth $d$ colors.
This has the effect that only parts of the neighborhood tree are being expanded, while other parts remain at the previous depth.
By employing the above procedure, we obtain colors $c^{(d_*)}$ with $c^{(d)}\sqsupseteq c^{(d_{*})}\sqsupseteq c^{(d+1)}$.
Note: The second refinement relation is not strict only if \emph{all} nodes associated with the selected depth $d$ colors are `pushing'.

\begin{figure*}[t]
		\includegraphics[width=\textwidth]{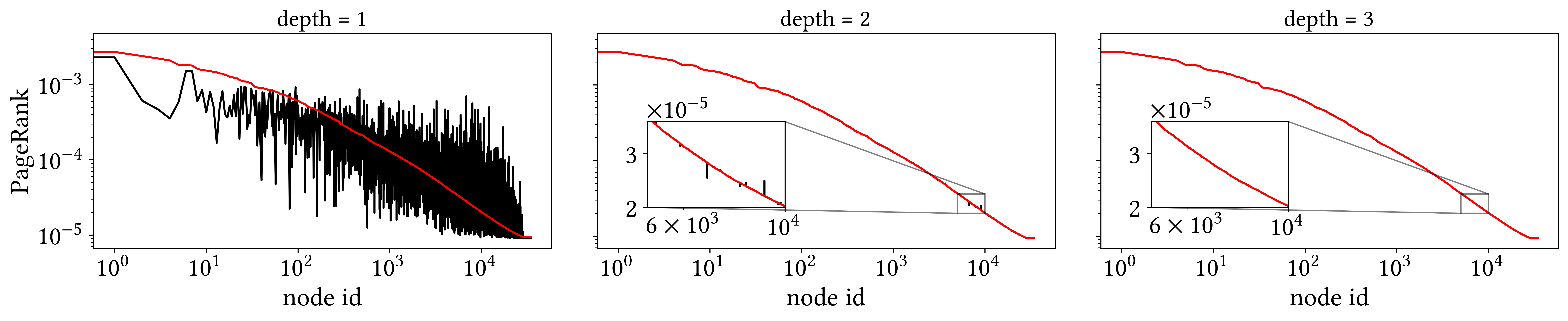}
		\vspace{-8mm}
		\caption{PageRank distribution of the \panicc{}$^{(d)}(\text{deg}_\text{out})$ model for different values of the depth $d$.
		We show the distribution of PageRank for the HepPh-network (red) and one sample from the fitted \panicc{$^{(d)}(\text{deg}_\text{out})$} model (black).
		We can see that for a depth of one (left) the PageRank score of the original network varies a lot compared with the PageRank of the sampled network. With increasing depth this error decreases and almost vanishes (max error $<10^{-16}$) for depth 3.
		}
	\label{fig::convergence}
\end{figure*}

\section{The role of the depth parameter $d$}
\label{sec:mathematical_properties}
\subsection{Maximum depth preserves centralities exactly}
The following theorem exemplifies how the (full) neighborhood information contained in the CR-induced node colors preserves many observables of a network. Specifically, we show how a range of centrality measures are preserved if appropriate choices are made for the color refinement procedure.
\begin{theorem}
\label{thm:EP_implies_centralities}
Let $G_1, G_2 \in \mnf^{\infty}_G(c^{(0)})$ be samples from the \panicc{} model with CR aggregating over the in-neighbors and let $A_1, A_2$ be their adjacency matrices. If two nodes $u,v \in V(G_1 \cup G_2)$ have the same color $c^{(\infty)}(u)=c^{(\infty)}(v)$, then:
    \begin{enumerate}
        \item $\Gamma_{\textbf{Katz}}(u) = \Gamma_{\textbf{Katz}}(v)$ 
        \item $\Gamma_{\textbf{EV}}(u) = \Gamma_{\textbf{EV}}(v)$.
        \item If $c^{(0)} \sqsubseteq \text{deg}_{\text{out}}$, then $\Gamma_{\textbf{PR}}(u) = \Gamma_{\textbf{PR}}(v)$. 
        \item If $c_{\text{in}}^{(\infty)}$ is computed for $A_i^\top A_i$, then $\Gamma_{\textbf{auth}}(u) = \Gamma_{\textbf{auth}}(v)$
        \item If instead CR aggregates over both in and out neighbors then $\Gamma_{\textbf{auth}}(u) = \Gamma_{\textbf{auth}}(v)$ and $  \Gamma_{\textbf{hub}}(u) = \Gamma_{\textbf{hub}}(v)$
    \end{enumerate}
\end{theorem}
We emphasize again that for unique centralities a corresponding \emph{primitive} matrix is not required, see our centrality definition in \cref{sec:preliminaries}.

\Cref{thm:EP_implies_centralities} shows, that the centrality of a node is completely determined by the node's stable color, i.e, the neighborhood tree encoded by the color implies the value of the centrality score --- even for completely unrelated graphs.
This means, that for sufficiently large $d$, many centrality measures of the original graph are preserved exactly in NeSt samples.

We remark that the HITS score factors in both $A$ and $A^\top$ which makes it necessary to regard both in- and outgoing neighbors in the CR computation (see \cref{eq:color_refinement_both}).
The statement for PageRank has previously been established in \cite{boldi2006graph}.

\noindent\textbf{The algebraic view of CR}
To prove this result, we do an established switch of our perspective from the combinatorial view of CR to an algebraic one.
Consider the following partition indicator matrix whose columns indicate the color class $\mathcal{C}_i = \{v \in V \mid c(v) = i\}$ a node belongs to:
\begin{equation}
\label{eq:indicator_matrix}
H^{(d)}_{i,j} = \mathbb{I}\left[i \in \mathcal{C}_j^{(d)}\right] =  \begin{cases}
1 & \text{ if } i \in \mathcal{C}^{(d)}_j\\
0 & \text{ else }
\end{cases}
\end{equation}

This indicator matrix $H$ can be used to count the number of neighbors of color class $i$ for node $u$ by simply multiplying $H$ with the adjacency matrix as follows:
\begin{equation}
\left(A^\top H^{(d)}\right)_{u,i} =  \mkern-12mu \sum_{v \in N_{\text{in}}(u)}\mkern-9mu \mathbb{I}{\left[v \in C^{(d-1)}_i\right]}
\label{eq:A_times_H}
\end{equation}
Once a stable coloring is reached, the partition described by $H$ is equitable which means each node from the same color class has the same number of colored neighbors.
Thus the rows of the matrix $AH^{(\infty)}$ are the same for all nodes of the same color class. This allows us the express this matrix as
\begin{equation}
\label{eq:algebraic_EP}
AH^{(\infty)} = H^{(\infty)} A^\pi,
\end{equation}
where $A^\pi= (H^\top H)^{-1}H^\top A H$ is the adjacency matrix of the so-called \emph{quotient graph}.
We omit the superscript $\infty$ when referring to the indicator matrix $H$ of the equitable partition.

The above considerations can be generalized for directed graphs, in which case the neighbourhood has to be replaced with either the out- or the in-neighbourhood:
$$
AH_{out} = H_{out}A^\pi_{out} \quad \text{ or } \quad A^\top H_{in} = H_{in} A^\pi_{in}
$$

\ifIEEE \begin{IEEEproof} \else \begin{proof} \fi \textit{(\cref{thm:EP_implies_centralities})}
The main idea to the proof is that each eigenvector of the $A^\pi$ matrix ($A^\pi w^\pi = \lambda w^\pi$) can be scaled up by multiplication with the partition indicator matrix $H$ (defined in \cref{eq:indicator_matrix} indicating the coarsest equitable partition) such that $Hw^\pi$ is an eigenvector of $A$ to the same eigenvalue:
$$
AHw^{\pi} = H A^\pi w^{\pi} = \lambda Hw^{\pi}
$$
In \cite{scholkemper2022blind} it is shown that the dominant eigenpair of $A$ is shared in this way if it is unique, which is the key insight toward proving the theorem, seeing as both $A_i$ have the same $H$ and $A_\pi$ by assumption. For (1) through (5) we always find that $\Gamma_{\textbf{X}} = H \Gamma_X^\pi$. Noticing that the blown-up vector has the same value for all nodes that are in the same WL-class (as indicated by the columns of $H$) yields that the nodes have the same centrality score.

(1) For Katz centrality, consider the definition:
$$
\begin{aligned}
\Gamma_{\textbf{Katz}} &\mskip-5mu=\mskip-6mu \sum_{k=1}^\infty\mskip-2mu a^k A^{ \top}\mskip-5mu \mathbbm{1}_n
\mskip-5mu=\mskip-6mu \sum_{k=1}^\infty\mskip-2mu a^k\mskip-2mu  A^\top\mskip-5mu H \mathbbm{1}_k 
\mskip-5mu=\mskip-5mu H\mskip-2mu\sum_{k=1}^\infty\mskip-2mu a^k \mskip-2mu A^\pi_{\text{in}} \mskip-2mu \mathbbm{1}_k
\mskip-5mu=\mskip-5mu H \Gamma_{\textbf{Katz}}^\pi
\end{aligned}
$$

(2) For eigenvector centrality, we have:
$$
\Gamma_{\textbf{EV}} \mskip-4mu=\mskip-11mu \lim_{m \rightarrow \infty}\mskip-4mu \frac{1}{m} \mskip-4mu \sum_{i = 0}^m \mskip-2mu (\lambda^{-1}\mskip-4mu A^\top)^i \mskip-2mu H \mathbbm{1} \mskip-4mu = \mskip-4mu H \mskip-9mu \lim_{m \rightarrow \infty} \mskip-4mu \frac{1}{m}  \mskip-4mu \sum_{i = 0}^m \mskip-2mu (\lambda^{-1} \mskip-2mu A^\pi_{\text{in}})^i \mathbbm{1}
$$

(3) For PageRank, notice that $c_{1,\text{in}}^{(\infty)}, c_{2,\text{in}}^{(\infty)} \sqsubseteq c^{(0)} \sqsubseteq d_{\text{out}}$ implies that $D_{\text{out}}^{-1}H = H D_{\text{out}}^\pi$.
Consider the page rank matrices $M_1, M_2$ for both graphs.
It holds that:
$$
\begin{aligned}
M_i H &= \alpha A_i^\top D_{\text{out}}^{-1} H + \frac{(1-\alpha)}{n}\mathbbm{1}_n\mathbbm{1}_n^\top H\\
&= H \left(\alpha A^\pi  D_{\text{out}}^{\pi} + \frac{(1-\alpha)}{n}\mathbbm{1}_k(|C_1|, ..., |C_k|)\right)
= HM^\pi
\end{aligned}
$$
As the graph $G_{M^\pi}$ that has $M^\pi$ as its adjacency matrix, is strongly connected and aperiodic, $M^\pi$ is primitive with perron (dominant) eigenvector $w_{\pi}$. This can then be scaled up to see that $Hw_{\pi}$ is shared between $M_1$ and $M_2$.

(4) Is similar to (2) but with $A$ being replaced by $A_i^\top A_i$.

(5) Let $G$ be a graph with adjacency matrix $A$. 
Let $H$ indicate the coarsest equitable partition found by the WL algorithm through aggregating both in- and out-neighbourhood, then: $A_1 H = HA^\pi = A_2 H$ as well as $A_1^\top H = HA^\pi_{\text{in}} = A_2^\top H$. We prove by induction that all iterations  $a_i^{(k)}, h^{(k)}$ are the same for both graphs and are of the form $a_i^{(k)} = Ha_\pi^{(k)}, h_i^{(k)} = Hh_\pi^{(k)}$. For the base case, $h_i^{(0)} = \mathbbm{1} = H \mathbbm{1}$.
For the induction step: 
$$a_i^{(k+1)} = \frac{A_i^\top Hh^{(k)}_\pi}{\lVert A_i^\top H h^{(k)}_\pi \rVert} = \frac{HA^\pi_{\text{in}} h^{(k)}_\pi}{\lVert HA^\pi_{\text{in}} h^{(k)}_\pi \rVert} $$
The same statement for $h_i^{(k)}$ can be shown accordingly. The final iterates have the form $a_i^{(\infty)} = H a^{(\infty)}_{\pi}, h_i^{(\infty)} = H h^{(\infty)}_{\pi}$ and the statement follows. 
\ifnotIEEE \end{proof} \else \end{IEEEproof} \fi

\subsection{Intermediate $d$ approximates centralities}

In the previous section, we showed that it is possible to preserve enough structure in samples from the \panicc{} model to keep centralities like PageRank invariant.
However for some purposes, it may suffice to only approximately preserve centrality scores while allowing for a richer set of possible network samples.
In the following, we thus consider cases in which preserving smaller neighborhood depths is already sufficient.
With PageRank as a running example, we show that convergence guarantees in the power iteration can be converted into approximation guarantees for samples drawn from the \panicc{} model. Remember the PageRank iteration for a graph $G$ with adjacency matrix $A$ is defined as 
$$\label{eq:PR_iteration}
    x^{(t+1)}(x^{(0)}) =  \alpha \bar A^\top D_{out}^{-1} x^{(t)}(x^{(0)})+\frac{1-\alpha}{n} \mathbbm{1}_n
$$
when starting from starting vector $x^{(0)}$ with no negative entries. Remember $\bar A$ is an augmented adjacency matrix. The explicit dependency on $x^{(0)}$ is omitted when it is clear from context.

The following result formalizes the consequences for samples from the \panicc{} model for PageRank centrality.

\begin{lemma}
\label{lemma:intermediate_colors}
Let $G$ be a graph and let $\tilde{G}$ be sampled from in-\panicc{$^d_G(\text{deg}_{\text{out}})$}. Then the first $d$ terms in the PageRank power iteration agree for $G$ and $\tilde G$:
$$x^{(t)} = \tilde{x}^{(t)}\,\; \forall t \leq d$$
Where $x^{(t)}$ is a power iteration for $G$ and $\tilde{x}^{(t)}$ be defined accordingly for $\tilde{G}$. Both power iterations start with $x^{(0)} = \tilde x^{(0)} = \text{const} \cdot \mathbbm{1}_n$.

\end{lemma}
The proof of this lemma also implies an equivalent result for $x^{(t+1)} = A^\top x^{(t)}$ used when computing eigenvector centrality, and can be adapted for any power iteration.
Notice that if $d$ is large enough such that the coloring is stable, the lemma implies statement (3) in \cref{thm:EP_implies_centralities}.

\noindent\textbf{The algebraic view for intermediate depth}
Towards a proof of \Cref{lemma:intermediate_colors}, we extend the previously established algebraic view of CR to intermediate colors.
That is, we derive statements akin to \cref{eq:algebraic_EP} for the intermediate colors.
As noted in~\cite{kersting2014power}, nodes that have the same rows in $A^\top H^{(d)}$ are in the same color class $\mathcal{C}^{(d+1)}_i$ and vice versa.
We can thus establish the following connection between consecutive iterations of CR:
\begin{equation}
\label{eq:intermediate_EP_statement}
A^\top H^{(d)} = H^{(d+1)} X_{d+1}^\pi
\end{equation}
where $X^\pi_{d+1}$ reflects the relationships between the colors at depth $d$ and $d+1$.
Now, because $G$ and $\tilde G$ sampled from in-\panicc{$^d(c_0)$} have identical colorings for $t \leq d$, thus they share the same $H^{(t)}$ and $X_{t}^\pi$ matrices. $A$ and  $\tilde A$ are thus related
$$
A^\top H^{(t-1)} = H^{(t)} X^\pi_{t} = \tilde{A}^\top H^{(t-1)}
$$
for all $t \leq d$.
These observations carry over to directed graphs.

\ifIEEE \begin{IEEEproof} \else \begin{proof} \fi
For a proof of \cref{lemma:intermediate_colors} %
first notice that $H_{\text{in}}^{(t)}$ and $D_{\text{out}}^{-1}$ are similarly related as the adjacency matrix, i.e. $D_{\text{out}}^{-1}H_{\text{in}}^{(t)} = H_{\text{in}}^{(t+1)}D_{t+1}^\pi$ for every $t$ because $D_{\text{out}}$ is encoded in the initial colors.
We now proceed by induction on $t$.
We show that $x^{(t)} = H_{\text{in}}^{(t)}x^{(t)}_\pi = \tilde x^{(t)}$, which holds for $t = 0$ and $x^{(0)} = \text{const} \cdot \mathbbm{1}_n$. 
Assuming the induction statement, it follows that:
\begin{subequations}\label{eq:intermediate_PR_EP_same_as_A}
\begin{align}
x^{(t+1)} &= \alpha A^\top D_{\text{out}}^{-1} x^{(t)} + \frac{(1-a)}{n} \mathbbm{1}_n\\
&= \alpha A^\top D_{\text{out}}^{-1} H_{in}^{(t)}x^{(t)}_\pi + \frac{(1-a)}{n} \mathbbm{1}_n\\
&= H^{(t+1)}_{\text{in}} \Big (\alpha X^\pi_{t+1} D_{t+1}^\pi x^{(t)}_\pi + \frac{(1-a)}{n} \mathbbm{1}_k\Big)
\end{align}
\end{subequations}
The last line is $x^{(t+1)} = H_{\text{in}}^{(t+1)}x^{(t+1)}_\pi$ which completes the induction. 
Repeating the same for $\tilde{x}^{(t+1)}$ concludes the proof.
\ifnotIEEE \end{proof} \else \end{IEEEproof} \fi

\noindent\textbf{Guaranteed similarity in PageRank}
We now extend the result that an original graph and samples from 
 in-\panicc{$^d_G(\text{deg}_{\text{out}})$} are constrained in their first power iterations (\Cref{lemma:intermediate_colors}) to the finding, that the final PageRank values of both graphs cannot be arbitrarily apart.
We achieve this by combining \Cref{lemma:intermediate_colors} with convergence guarantees of the PageRank power iteration. We can thus be sure that two samples from the \panicc{} model are bound to have centralities that are no further apart than the following:

\begin{lemma}
\label{lemma:pr_convergence}
Let $x$ and $\tilde x$ be the PageRank vectors of two graphs sampled from in-\panicc{$^d_G(\text{deg}_{\text{out}})$}. Let $\alpha$ be the PageRank damping factor used.
It holds that:
$$\lVert x-\tilde{x} \rVert_1 \leq 2 \alpha^{d+1}$$
\end{lemma}
Informally, the probability mass of the PageRank vector, for which the iterates of the original and the synthetic network do not yet agree upon, is of magnitude at most $\alpha^{d+1}$.
Hence, the final PageRank vectors are at most twice this magnitude apart.
\ifIEEE \begin{IEEEproof} \else \begin{proof} \fi

We use Theorem 6.1 in \cite{bianchini2005inside} that states:
\begin{equation}
  \lVert x - x^{(i)}(y) \rVert_1 \leq \alpha^j \lVert x - x^{(i-j)}(y) \rVert_1
  \label{eqn::worst_case}
  \end{equation}
for any non negative $y$ with  $\lVert y \rVert_1\leq 1$.
Similar to \cite{gleich2015pagerank} we find the relationship $x^{(i+1)}(0) = x^{(i)}(\tfrac{1-\alpha}{n}\mathbbm{1}_n)$. From this we get:
\begin{align*}
 \norm{x - x^{(i)} (\tfrac{1-\alpha}{n}\mathbbm{1}_n ) } &= \norm{x - x^{(i+1)}(0) } \leq \alpha^{i+1} \norm{x - x^{(0)}(0) }\\ &= \alpha^{i+1} \norm{x} = \alpha^{i+1}   
\end{align*}

We conclude the proof using the triangle inequality (1), \Cref{lemma:intermediate_colors} (2), and the directly above relationship (3). 
$$
\begin{aligned}
\lVert x - \tilde{x}\rVert_1 & \overset{(1)}\leq \lVert x - \tilde{x}^{(d)} \rVert_1 + \lVert \tilde{x}^{(d)} - \tilde{x} \rVert_1\\
& \overset{(2)}= \lVert x - x^{(d)} \rVert_1 + \lVert \tilde{x} - \tilde{x}^{(d)} \rVert_1 \overset{(3)}\leq \alpha^{d+1} + \alpha^{d+1} 
\end{aligned}
$$
\ifnotIEEE \end{proof} \else \end{IEEEproof} \fi

These theoretical considerations provide only worst-case bounds
which are typically not tight for many (real-world) graphs --- as one can see by considering the case that for regular graphs an equitable partition can already be reached at $d=1$.
Then \Cref{lemma:pr_convergence} yields a bound of $\approx 1.4$ (using typical $\alpha$=0.85), while we know from \Cref{thm:EP_implies_centralities} that the actual difference is $0$.
The next bound provides better guarantees in these cases.

\noindent\textbf{Convergence of iteration implies similarity in PageRank}
In many real-world networks, the PageRank iteration converges faster than the worst case considered in \cref{eqn::worst_case}.
We thus establish a second bound which relates the convergence in one network to a guaranteed similarity in PageRank.
\begin{corollary}
\label{cor:pr_convergence}
With assumptions as in \Cref{lemma:intermediate_colors,lemma:pr_convergence}:
$$
\begin{aligned}
\lVert x - \tilde{x}\rVert_1 \leq \frac{2}{1-\alpha} \lVert x^{(k-1)} - x^{(k)}\rVert_1
\end{aligned}
$$
\end{corollary}
Colloquially speaking, \Cref{cor:pr_convergence} states that if the PageRank iterations in the original network converge quickly, then the PageRank vectors of synthetic and original are not far apart.

\ifIEEE \begin{IEEEproof} \else \begin{proof} \fi
Mirroring the reasoning of \cite{gleich2015pagerank}, we have:
\begin{align*}
\lVert x^{(k-1)} - x\rVert_1 &\leq \frac{1}{1-\alpha} \lVert x^{(k-1)} - x^{(k)}\rVert_1
\end{align*}
Now, we have $x^{(t)} = \tilde{x}^{(t)}$ for $t \leq k$ (\cref{lemma:intermediate_colors}).
Therefore:
\begingroup
\small
\setlength{\thickmuskip}{0.5mu}
\setlength{\medmuskip}{0.5mu}
\setlength{\thinmuskip}{0.5mu}
\begin{align*}
&\lVert x - \tilde{x}\rVert_1 \leq \lVert x - \tilde{x}^{(k)} \rVert_1 + \lVert \tilde{x}^{(k)} - \tilde{x} \rVert_1
= \lVert x - x^{(k)} \rVert_1 + \lVert \tilde{x} - \tilde{x}^{(k)} \rVert_1\\
&= \tfrac{1}{1-a} \left(\lVert x^{ (k-1)} - x^{(k)}\rVert_1 + \lVert \tilde{x}^{(k-1)} - \tilde{x}^{(k)}\rVert_1 \right)
= \tfrac{2}{1-a} \lVert x^{(k-1)} - x^{(k)}\rVert_1
\end{align*}
\ifnotIEEE \end{proof} \else \end{IEEEproof} \fi
\endgroup

\label{sec:main_proof}

\begin{figure}[htbp]
\captionsetup[subfigure]{aboveskip=0.5pt,belowskip=3pt}

	\includegraphics[width=1\columnwidth]{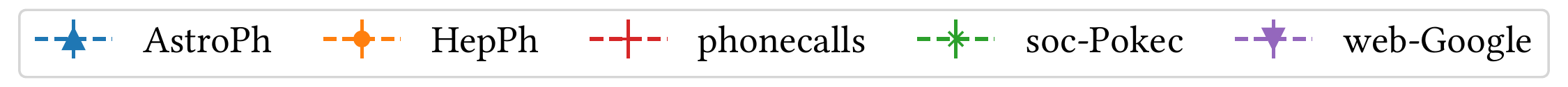}

\begin{subfigure}[b]{0.487\columnwidth}
	\includegraphics[width=1\columnwidth] {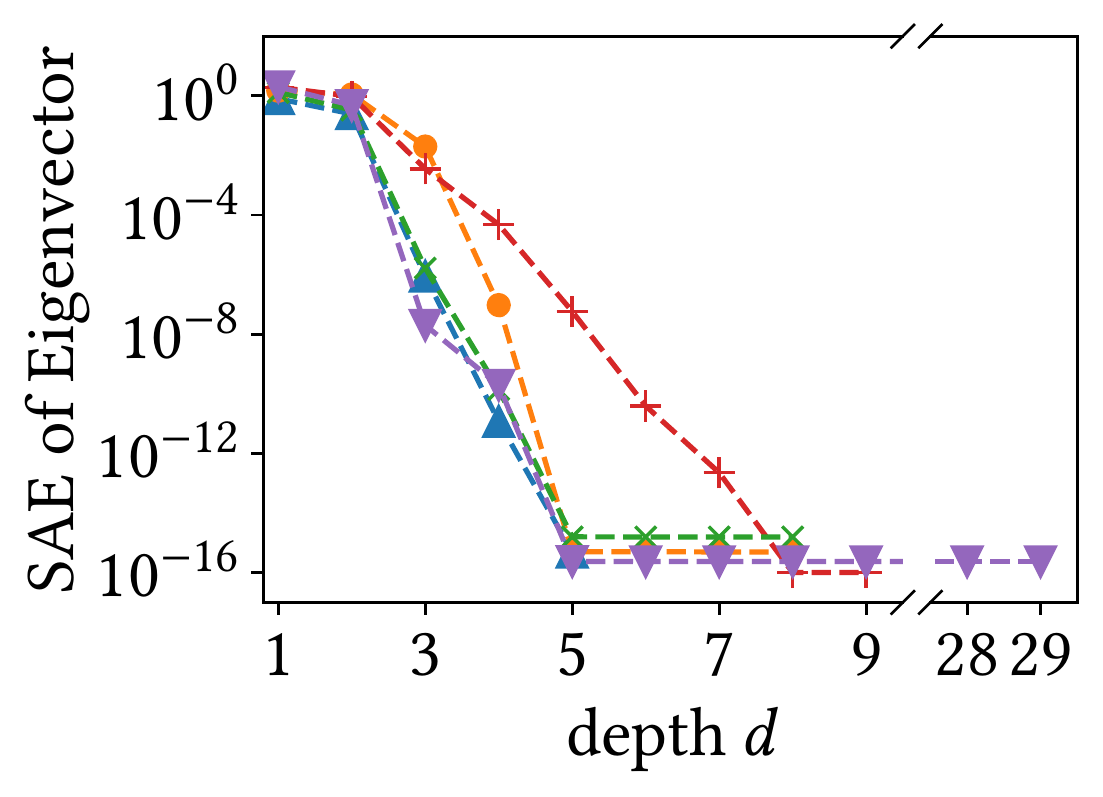}
        \caption{$\Gamma_{\textbf{EV}}$ of in-\panicc{}(uniform)}
	\label{fig::eigenvector}
\end{subfigure}
\begin{subfigure}[b]{0.48\columnwidth}
	\includegraphics[width=1\columnwidth] {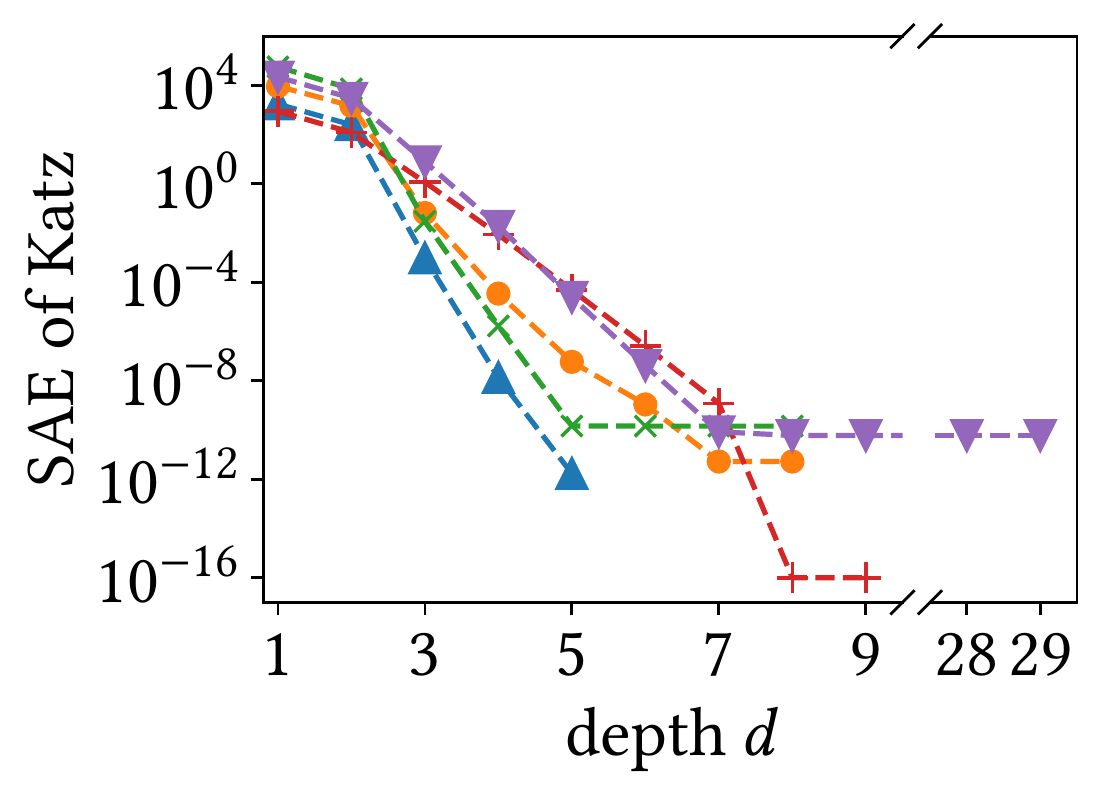}
		\caption{$\Gamma_{\textbf{Katz}}$ of in-\panicc{}(uniform)}
	\label{fig::katz}
\end{subfigure}
\begin{subfigure}[b]{0.48\columnwidth}
	\includegraphics[width=1\columnwidth]{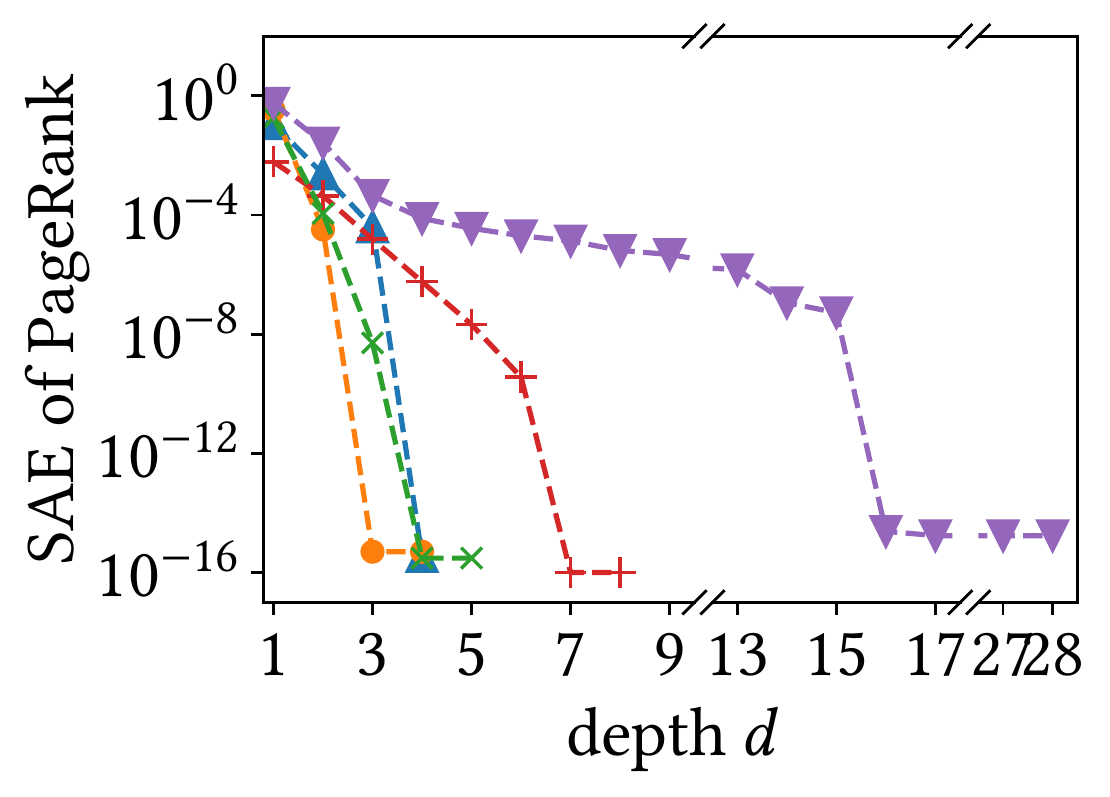}
		\caption{$\Gamma_{\textbf{PR}}$ of in-\panicc{}($\text{deg}_{\text{out}}$)}
	\label{fig::pagerank}
\end{subfigure}
\begin{subfigure}[b]{0.48\columnwidth}
	\includegraphics[width=1\columnwidth]{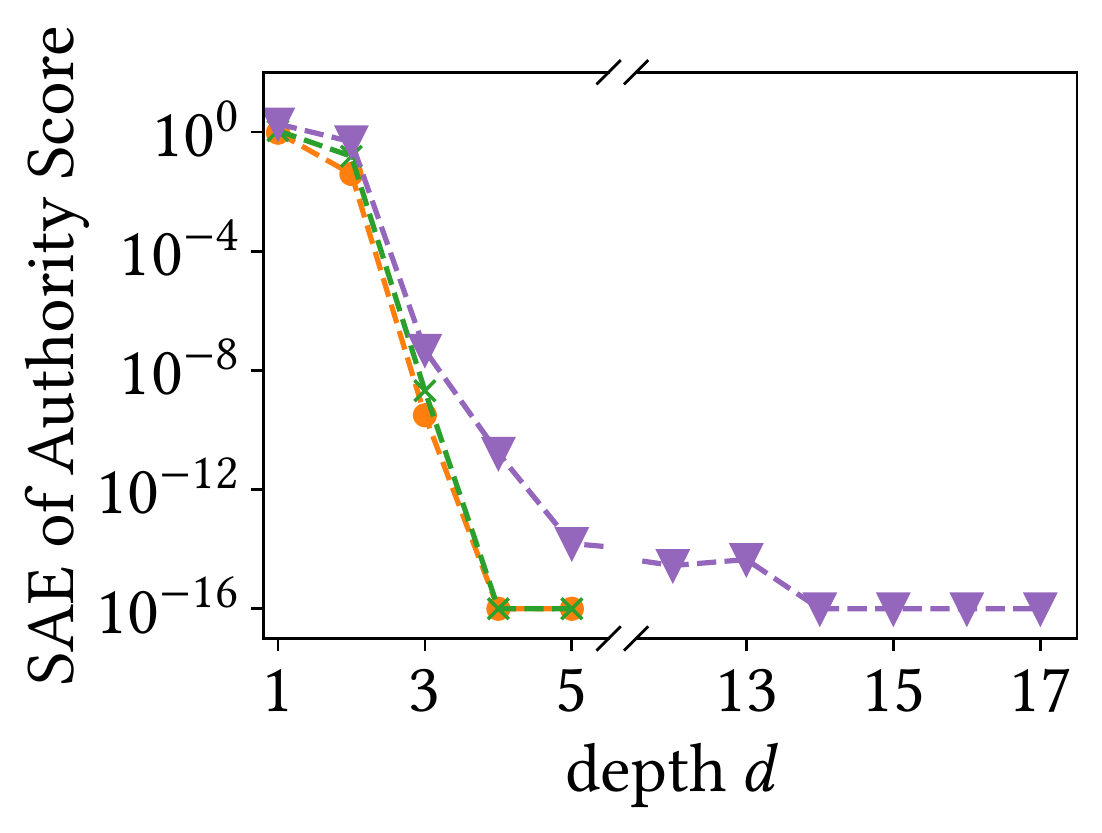}
		\caption{$\Gamma_{\textbf{Auth}}$ of both-\panicc{}(uniform)}
	\label{fig::auth}
\end{subfigure}

\caption{Centralities are better and better preserved with increasing depth. We show the sum absolute error (SAE) for a sampled network in relation to the original network.
Points are medians with 16\%/84\% quantile error bars for 100 samples. Values are capped below by $10^{-16}$. Legend is on top.
From left to right increasingly deeper neighborhood trees are preserved which leads to centrality measures being better and better preserved.}
\label{Fig:summary}
\vspace{-0.5cm}
\end{figure}

\begin{figure}[tbp]
		\includegraphics[width=0.95\columnwidth]{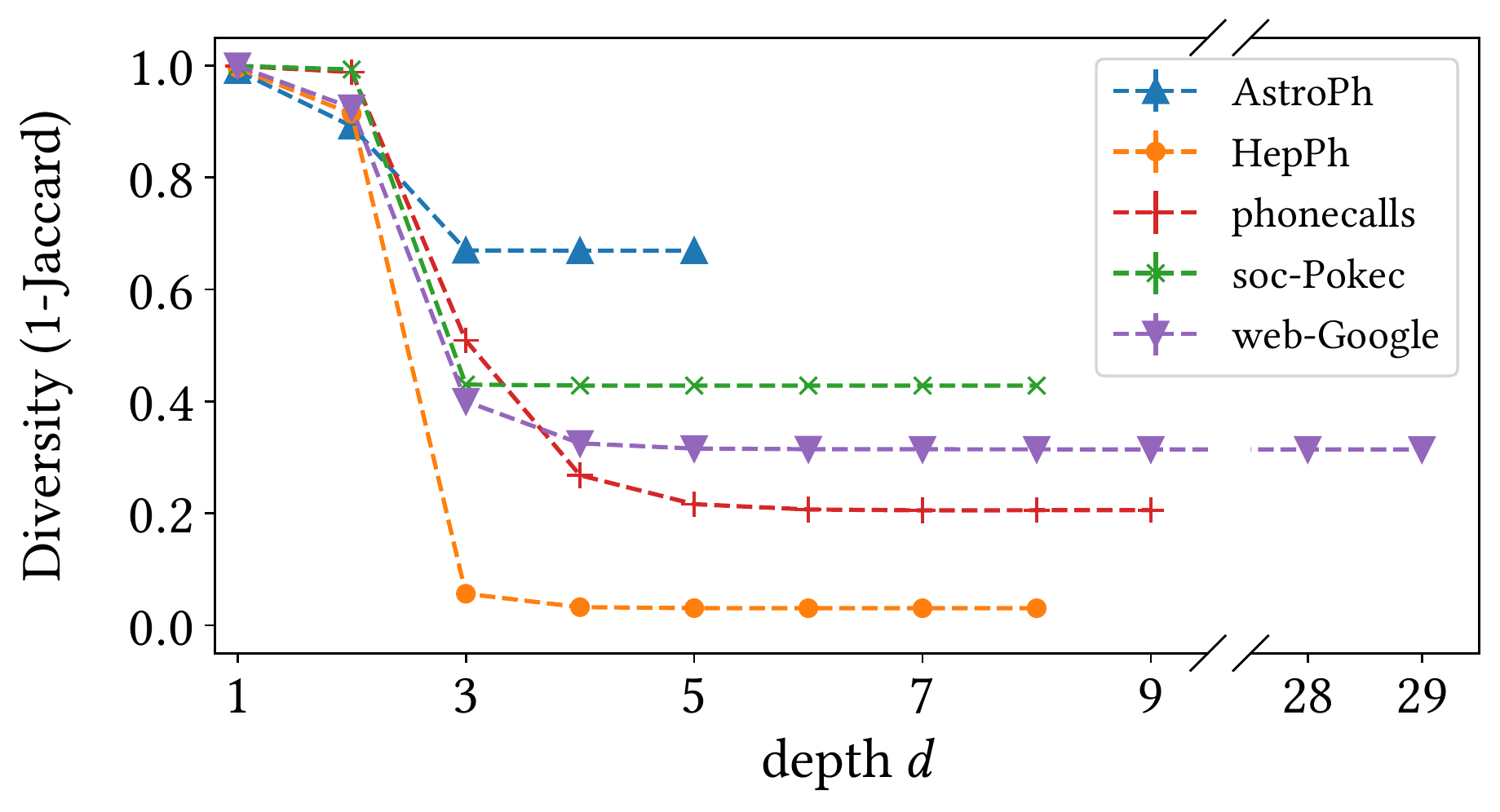}
		\vspace{-3mm}
		\caption{Diversity in \panicc{}(const) samples.
		We measure Diversity as one minus the Jaccard-Similarity of the edge sets of the original and sampled network.
		For a diversity of 1 the original and sampled network share no edges, for a diversity of 0 the networks agree perfectly.
		We see that for most networks, except for the HepPh network, there is still quite some diversity in the network samples. That is despite those networks better and better preserving centrality measures compare \cref{Fig:summary}.
		}
	\label{fig::diversity}
\end{figure}

\begin{figure}[bt]
    \centering
    \small
    \begin{tabular}{c|c|c|c|c} 
 
 Name & directed & EV & \#nodes & \#edges  \\ 
 \hline
 Karate & \xmark & \cmark & 34 & 78 \\ 
 \hline
 AstroPh & \xmark & \cmark & 18.772 & 198.110  \\ 
 \hline
 phonecalls & \xmark  & \cmark  & 36.595 & 45.680 \\ 
 \hline
 HepPh & \cmark & \cmark & 34.546 & 421.578 \\ 
 \hline
 web-Google & \cmark & \cmark & 875.713 & 5.105.039 \\ 
 \hline
 soc-Pokec & \cmark & \cmark & 1.632.803 & 30.622.564 \\

\end{tabular}
    \caption{Statistics of the real world networks used. EV indicates whether the dominant eigenvector is unique.}
    \label{fig:table}
\end{figure}

\section{Empirical Illustration}
\label{sec:empirical}

We augment our theoretical considerations with an empirical evaluation on a variety of real-world networks (both directed and undirected) from different domains. 
We include the citation network HepPh, the web graph web-Google, the social network soc-Pokec, the collaboration network AstroPh (all previous are from \cite{snapnets}), and a network of phonecalls~\cite{phonecalls}.
For details on the networks see \Cref{fig:table}.

We compute the CR colors using the indicated aggregation strategy and starting colors.
We then generate samples from the \panicc{} model for each of the centralities, and compute the centrality measures for the sampled networks.
Centralities are computed using a suitable iterative method until the SAE between subsequent iterations falls below a convergence threshold of $10^{-15}$.
As starting vector we chose the normalized (1-norm) vector for PageRank and the normal all-ones vector for other centralities.

We exemplify detailed convergence results for the PageRank distribution on the HepPh-network in~\Cref{fig::convergence}.
To increase visibility, nodes are sorted in descending order by their original PageRank score.
In the left plot at depth $d=1$ (one step of CR starting from out-degree colors), we see that there is quite a large difference in PageRank.
While the maximum absolute error (MAE) is below $0.002$, the relative error can be an order of magnitude.
In the middle plot ($d=2$), the sampled network closely approximates the true PageRank with the MAE dropping three orders of magnitude.
In the rightmost plot ($d=3$) the sample reflects the PageRank almost perfectly (MAE drops by a factor of $10^{-10}$). This shows, that meso scale properties are indeed relevant well recover the PageRank centrality in network samples.

Convergence results for other networks are summarized in \Cref{Fig:summary}.
Here we no longer show the individual distributions but the sum of absolute error (SAE) of the centrality of a sample in comparison to the original network averaged over $100$ samples.
The first two plots (\Cref{fig::eigenvector,fig::katz}) show eigenvector centrality and Katz centrality for the \panicc{} model which preserves in-neighborhood trees starting from uniform colors.
The third plot shows PageRank for the \panicc{} which preserves in-neighborhood trees starting from a coloring that reflects the out degree.
The last plot (\Cref{fig::auth}) shows the Authorities (HITS) score for the \panicc{} which preserves both the in- and out-neighborhood trees starting from uniform colors.

As expected from out theoretical consideration, for sufficient depth $d$ the sampled networks and original network are identical in their centrality scores.
We further find that for smaller $d$ networks already start to become more similar to the original network in terms of their centralities, which we proved for PageRank only.
We finally note, to achieve significant reduction in SAE preserving neighborhood trees of depth of at least three seems necessary.

\begin{figure}[btp]
\includegraphics[width=1\columnwidth] {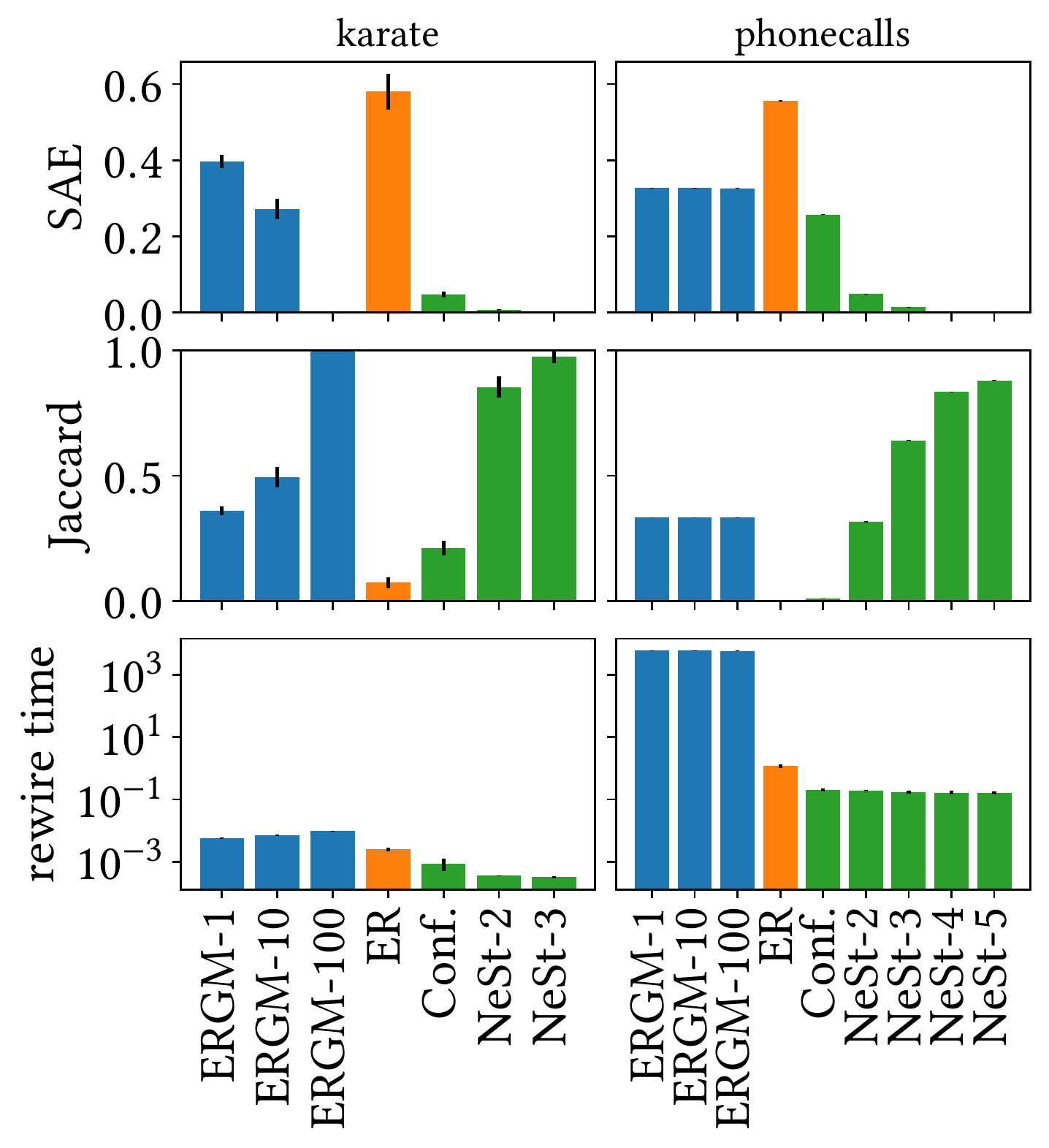}

\caption{Comparison of the NeSt model with other random graph models. The left/right column correspond to the Karate Club/ phonecalls network with 34/ 36k nodes. The rows show sum absolute error (SAE) of the PageRank (top), Jaccard similarity (mid) and rewire time in seconds on a logarithmic scale (bottom). We used algorithm~\ref{alg:sampling2} with $2 |E|$ steps. Lower is better on all scales. Besides NeSt we include Exponential Random Graph Models (ERGM), the Configuration model and the Erd\H{o}s-R\'enyi (ER) network. On the small network (left) both ERGM and NeSt allow a trade-off of similarity for SAE but ERGM has poor runtime. On the larger network (right) we still have the same tradeoff while NeSt shows fast runtimes similar to Config./ER, but ERGMs are no longer feasable (runtime). }
\label{fig::baselines}
\end{figure}

Better preserved neighborhood structure usually means diminished diversity in network samples, we show the extend of this in \Cref{fig::diversity}. 
We see, that for depth $1$ and $2$ the diversity is usually not diminished much, while for higher depth the diversity goes down for all networks settling at some final diversity. This can be pretty high for networks like AstroPh or low as in the case of HepPh.

Finally we compare the NeSt model with other random graph models for the task of generating networks with similar PageRank as an original network.
We compare the different models on three different scales (compare \Cref{fig::baselines}) that highlight different dimensions of the task: SAE is the main objective measuring similarity in terms of their PageRank, Jaccard similarity measures similarity of the edge sets of the sample to the original network and is a proxy for sample diversity and lastly rewire time measures how long it takes for one sample to be generated.
Lower is better on all scales.
Shown are averages and their standard deviation obtained for 100 runs of the algorithms with different seeds. To have a fairer comparison of runtime, algorithm 2 was used to sample form NeSt.

The relatively fast ER model offers a large variety of network samples (low Jaccard) but does not well reflect the PageRank (high SAE). Comparing runtime of ER to ERGM or NeSt should be avoided because it is not an apples to apples comparison but included for completeness.
The ERGM used (details see appendix) allows to trade Jaccard similarity for similarity in PageRank on the karate network but fails to do so on the phonecalls network.
Unfortunately, sampling from the ERGM is already impractical slow on the phonecalls network.
The ERGM is slow because the PageRank needs to be recalculated for each step.

The NeSt model and the known Configuration model (or alternatively NeSt$^{d=1}(\text{const})$)
allow to trade SAE for Jaccard on both the small network and the larger phonecalls network while maintaining a reasonable sampling time.
On the karate network, sampling NeSt models seems to become faster with increasing depth which goes against first intuition. But there are two effects at play: Firstly with increasing depth parts of the edges are frozen and are not considered for rewiring.
Secondly, rewiring many smaller graphs (higher depth) is faster than rewiring fewer bigger graphs (lower depth).

\noindent\textbf{Implementation details}
For the CR-algorithm we use an implementation suggested by Kersting et al.~\cite{kersting2014power}.
The code is available at \url{https://github.com/SomeAuthor123/NestModel}.

\clearpage
\section{Discussion}
\noindent\textbf{Preserving centralities}
As shown in \Cref{sec:main_proof}, the \panicc{} models not only (approximately) preserve centrality measures, but also the first iterates (up to depth $d$) of a corresponding power iteration.
Because we keep these early iterates invariant as well, \panicc{} is not sampling from \emph{all} networks with the same centrality as the original one.
Stated differently, there can be networks with different neighborhood trees that have the same centralities.
Thus our model is not an attempt to exactly preserve centrality scores.
In fact, we believe that the ability to maintain the neighborhood tree structure is more meaningful, as the network structure itself is the fundamental data we observe, whereas network statistics such as centrality measures are \emph{derived} from the network structure.

\noindent\textbf{Limiting the number of colors}
The CR algorithm can lead to color classes that contain just a single node, e.g., if there is a node with a unique degree.
As a consequence, the connectivity pattern to that node is frozen in \panicc{} of larger depth.
However in applications, it might be undesirable to distinguish nodes with $1000$ and $1001$ neighbors.
Consequently it might not be worth to preserve neighborhood trees exactly but rather approximately. This may be achieved by employing a clustering algorithm rather than a hash function when deciding for new colors.

\section{Conclusion}
In this paper, we have introduced \panicc{} models which enable the creation of synthetic networks that preserve the neighborhood tree structure of nodes in a given network up to a specified depth.
This allows to adjust the amount of structural information preserved in samples from null models to the task at hand.
\panicc{} models thus represent a versatile generalization of existing configuration models which can only preserve the degrees of nodes.
We demonstrate that \panicc{} models are efficient to fit through the Color Refinement Algorithm and easy to sample from, even for large networks.

While we illustrate the utility of preserving neighborhood structure by applying \panicc{} models for preserving centralities, the capabilities of the \panicc{} model extend to the preservation of many eigenvectors (e.g. the main eigenvalues \cite{rowlinson2007main} are preserved for sufficient $d$).
This could open up \panicc{} models as possible candidate null models for other spectral properties of a given network.
In fact, such spectral properties are important for a range of dynamical processes on networks such as (cluster) synchronization~\cite{schaub2016graph}, consensus dynamics~\cite{o2013observability}, or questions related to controllability~\cite{cardoso2007laplacian}.

Further, an interesting connection between \panicc{} models to message passing Graph Neural Networks (GNNs) exists: It has been shown that GNNs with $d$ layers and uniform node initialization are no more expressive than the first $d$ iterations of the CR-algorithm~\cite{morris2019weisfeiler,xu2018powerful, barcelo2020logical}.
As all samples from the \panicc{}$^{(d)}$ model are identically colored during the first $d$ CR iterations, they are thus indistinguishable by those GNNs. 
This opens up potential applications of the \panicc{} model and its variants for GNNs, e.g., to create (difficult) benchmark data sets.
In summary, we believe that \panicc{} models can provide a versatile and efficient instrument for network scientists that aim to produce network null models that conserve the larger neighbourhood structure of networks.

\bibliographystyle{ACM-Reference-Format}
\bibliography{references}

\clearpage
\newpage
\section{Appendix}

\label{sec:appendix}
\subsection{Proof of \cref{thm:sampling-correct}}

Throughout this proof, we prove the statement for a sub-graph $g_{C_i, C_j}$, as that directly implies the statement for the whole graph. Toward this end, we must prove that the Markov chain used to sample the graphs is connected, aperiodic and doubly stochastic \cite{fosdick2018configuring}. We start with the proof of connectedness. 
Let $G$ be a graph, $c^{(0)}$ the initial colors and $d \in \mathbbm{N}^+$. Further, let $O(G,d,c^{(0)})$ be the set of possible outputs of algorithm \ref{alg:sampling} with these input parameters.

\begin{claim}
$$
O(G,d,c^{(0)}) = \mnf^d_G(c^{(0)})
$$
\end{claim}

\ifIEEE \begin{IEEEproof} \else \begin{proof} \fi
''$\subseteq$'' The sampling procedure only edits edges in $g_{c^{(d-1)}_i, c^{(d-1)}_j}$. Consider a single edge flip involving $u_1, u_2, v_1, v_2$ as in algorithm \ref{alg:sampling} with  $c^{(d-1)}(u_1) = c^{(d-1)}(u_2)$ and $c^{(d-1)}(v_1) = c^{(d-1)}(v_2)$ by definition.
We prove the most restrictive case where colors are aggregated in both directions as in \cref{eq:color_refinement_both} and edges are directed from $c^{(d-1)}_i$ to $c^{(d-1)}_j$.   

Let 
$M_{\text{X}}^{(d-1)}(v) = \lmulti c^{(d)}(x) \mid x \in N_{\text{X}}(v) \rmulti$ be the multi-set of colors of neighbouring nodes for $X \in \{\text{in}, \text{out}\}$. Then \cref{eq:color_refinement_both} can be rewritten as:
$$
\begin{aligned}
c^{(d)}(u_1) = &\operatorname{hash}\mskip-5mu\Big(c^{(d-1)}(u_1), M_{\text{in}}(u_1), \\
& \left(M_{\text{out}}(u_1) \backslash \lmulti c^{(d-1)}(v_1)\rmulti\right) \cup \lmulti c^{(d-1)}(v_1)\rmulti  \Big)
\end{aligned}
$$
We prove by induction that colors $\tilde{c}^{(t)}$ in the new graph $\tilde{G}$ remain unchanged for all involved nodes. The base case for the initial colors holds per definition. For the induction step, assume the statement holds for $t$ and consider node $u_1$:
$$
\begin{aligned}
&\tilde{c}^{(t+1)}(u_1) \\
&=\mskip-5mu \operatorname{hash}\mskip-5mu\Big(\mskip-3mu \tilde{c}^{(t)}\mskip-2mu(u_1), \tilde{M}_{\text{in}}(u_1), 
 \tilde{M}_{\text{out}}(u_1) \backslash \lmulti\tilde{c}^{(t)}(v_1)\rmulti \cup \lmulti\tilde{c}^{(t)}(v_2)\rmulti \mskip-5mu \Big)\\
&=\mskip-5mu \operatorname{hash}\mskip-5mu\Big(\mskip-3mu c^{(t)}\mskip-2mu(u_1), M_{\text{in}}(u_1), 
 M_{\text{out}}(u_1) \backslash \lmulti c^{(t)}(v_1)\rmulti \cup
 \lmulti c^{(t)}(v_2)\rmulti  \mskip-5mu\Big)\\
&=\mskip-5mu \operatorname{hash}\mskip-5mu\Big(\mskip-3mu c^{(t)}\mskip-2mu(u_1), M_{\text{in}}(u_1), 
 M_{\text{out}}(u_1) \backslash \lmulti c^{(t)}(v_1)\rmulti \cup \lmulti c^{(t)}(v_1)\rmulti \mskip-5mu \Big)\\
 &= \ c^{(t+1)}(u_1) 
\end{aligned}
$$
Here the first equality is the result of the induction statement and the second equality holds because, in the original graph, $c^{(d)}(v_1) = c^{(d)}(v_2)$ implies that $c^{(t)}(v_1) = c^{(t)}(v_2)$ for $t \leq d$.
The same reasoning applies to the remaining nodes $u_2, v_1, v_2$. 

''$\supseteq$'' The backward direction is somewhat less intuitive. We show that any graph $G'$ with the same CR colors of depth $d$ can be reached by a sequence of at most $\frac{|D|}{2}-1$ edits, where $D = (E(G) \cup E(G')) \backslash (E(G) \cap E(G'))$ is the set of edges $G$ and $G'$ don't agree upon.
We again concern ourselves with the subgraphs $g_{C_i, C_j}$ as using edge flips within these sub-graphs is sufficient to convert $G$ into $G'$, since if all edges in all subgraphs agree then $G$ and $G'$ are the same. Let $g, g'$ be the subgraphs to the same pair of colors for $G, G'$ respectively.

Base case: $|D| = 4$.  Let $e_1 \neq e_2 \in D \cap E(g)$, $e_1' \neq e_2' \in D \cap E(g')$ and $e_1 = (u_1, v_1), e_2 = (u_2, v_2)$. Then it must be that $e_1' = (u_1, v_2), e_2'= (u_2, v_1)$ (apart from renaming).
As $G$ and $G'$ are both in $\mnf^d_G(c^{(0)})$ we have that $c^{(d)}_G(u_1) = c^{(d)}_{G'}(u_1)$, which implies that $\text{deg}^{\text{out}}_g(u_1) = \text{deg}^{\text{out}}_{g'}(u_1)$. As $G$ and $G'$ agree on all other edges, this means $e_1' = (u_1, \cdot)$ or $e_2' = (u_1, \cdot)$. Similar reasoning applied to $u_2$ yields $e_1' = (u_1, \cdot)$ and $e_2' = (u_2, \cdot)$. Repeating the same for the in-degree of the $v_i$'s and noting that $(u_1, v_1) \notin E(G')$ yields the statement.

Induction step: $n \rightarrow n-1$. Let $|D| \cap (E(g) \cup E(g')) = 2n$. Let $e_1 = (u_1, v_1) \in D \cap E(g)$ be an edge that $g$ and $g'$ do not agree on. Since the degree in the subgraphs must be the same(see base case), there must be at least one edge $e_1' = (u_1, v_1') \in D \cap E(g')$ that $g$ and $g'$ also do not agree on. Since the in-degree of $v_1'$ must also obey this, there exists an edge $e_2 = (u_2, v_1') \in D \cap E(g)$. In the case that $g = g_{C_i, C_j}$ for $C_i \neq C_j$, we have that all 4 nodes mentioned here are distinct. It could however be the case, that the edge $(u_2, v_1) \in E(g)$, which would prevent the edge flip. This implies that $\text{deg}^{\text{in}}_g(v_1) > \text{deg}^{\text{in}}_{g'}(v_1)$ and there exists another node $x$ and an edge $(x, v_1) \in D \cap E(g')$. If this edge flip is also prevented by a edge, then the in degree of $v_1'$ is again higher and we find another node. Since this cannot continue forever, as the graph is finite, we eventually find a node that we can use as $u_2$, i.e. where the edge flip is allowed. Performing an edge flip on $e_1, e_2$ yields the new edges $(u_1, v_1'), (u_2, v_1)$. Thus after this flip, $e_1$ has been transformed into $e_1'$, so the graphs now agree on one more edge compared to before. It is possible that $(u_2, v_1) \in D \cap E(g')$, in which case the edge flip relieved two disagreements simultaneously. In any case, $|D^*| \leq 2(n-1)$, where $D^*$ is the disagreement set between the new graph created from $G$ by the edge flip and $G'$.

However, in the case that $g$ is not bipartite, it may be the case, that $u_2 = v_1$. If we can choose any other configuration to get 4 distinct nodes, then we do so and treat it like the previous case. If we cannot, then there is no $(v_1, x) \in E(g')$ and no $(x, v_1') \in D \cap E(g)$ for $x \neq u_2$. Then, $u_1, v_1, v_1'$ constitute a directed triangle in $g$ and $g'$ with opposite directions. For this corner case, we need a new move, that turns one into the other, decreasing $|D|$ by 6, i.e. $|D^*| \leq 2(n-3)$.
\ifnotIEEE \end{proof} \else \end{IEEEproof} \fi

We have now shown that all graphs that have the same CR colors of depth $d$ are a possible output of algorithm \ref{alg:sampling}. Or in other words, the markov chain is connected. It is also aperiodic as choosing $e_1 = e_2$ is allowed and introduces a self loop.
Finally, the Markov chain to sample the subgraph $g_{C_i, C_j}$ for any fixed $C_i, C_j$ is row stochastic, since the number of edge pair choices (= number of possible edits) is the same for all allowed subgraphs. Note that some of the possible edits may be invalid and make up a self-loop. Finally, every edit can also be reverted, meaning the markov chain is symmetric and thus doubly stochastic. 
\subsection{The ERGM used}
In this work we used an ERGM with probabilities
$$
p(\tilde G) \propto \exp(-10 \cdot \theta |\Gamma_{PR}(\tilde G)-\Gamma_{PR}(G)|)
$$
The parameter $\theta$ controls how strongly graphs should resemble the PageRank score of the original graph $G$. We sample using the dyad flip Markov chain, i.e.from a current state a new graph is proposed which is obtained by flipping one dyad uniform at random from all dyads.

\end{document}